# BPP: Large Graph Storage for Efficient Disk-Based Processing

Kamran Najeebullah, Kifayat Ullah Khan, Muhammad Waqas Nawaz, Young-Koo Lee

Department of Computer Engineering, Kyung Hee University,
Yongin-si, Gyeonggi-do, Korea
{kamran, Kifayat, waqas}@dke.khu.ac.kr, yklee@khu.ac.kr

**Abstract.** Processing very large graphs like social networks, biological and chemical compounds is a challenging task. Distributed graph processing systems process the billion-scale graphs efficiently but incur overheads of efficient partitioning and distribution of the graph over a cluster of nodes. Distributed processing also requires cluster management and fault tolerance. In order to overcome these problems GraphChi was proposed recently. GraphChi significantly outperformed all the representative distributed processing frameworks. Still, we observe that GraphChi incurs some serious degradation in performance due to 1) high number of non-sequential I/Os for processing every chunk of graph; and 2) lack of true parallelism to process the graph. In this paper we propose a simple yet powerful engine BiShard Parallel Processor (BPP) to efficiently process billions-scale graphs on a single PC. We extend the storage structure proposed by GraphChi and introduce a new processing model called BiShard Parallel (BP). BP enables full CPU parallelism for processing the graph and significantly reduces the number of non-sequential I/Os required to process every chunk of the graph. Our experiments on real large graphs show that our solution significantly outperforms GraphChi.

## 1 Introduction

Graph processing has been a popular research area in the last decade and a lot of research has been targeted at the most common graph processing algorithms such as shortest path and some variations of clustering and page rank. Algorithms like connected components and minimum cut also have their own vital value. Graphs like social networks, biological and chemical compounds are difficult to process because of their massive size. With the growing size of the graph datasets, processing graphs has become more challenging.

GraphChi [1] processes very large graphs on a single PC using an asynchronous computation model. It introduced a naive processing model, Parallel Sliding Window (PSW). PSW performs computation in execution intervals. An execution interval consist of three steps 1) load a subgraph of the input graph into memory; 2) process all the vertices of the subgraph in parallel and modify the values associated with the vertices and their incident edges; and 3) write the updates back to the disk. PSW passes information between the vertices using out-edges. GraphChi significantly outperformed the distributed processing systems on per node basis.

We observe PSW inherits two serious bottlenecks. First, in every execution interval, PSW needs to perform non-sequential reads proportional to the number of the subgraphs. For a very large graph, in every execution interval the number of non-sequential reads will be significantly high. Second, for edges having both endpoints inside the same interval, PSW needs to avoid the race conditions between the two endpoints of the edge. As GraphChi maintains only one copy of the edges, both of the endpoints may end up accessing the edge at the same time. To avoid race conditions, PSW marks such edges as critical and processes the endpoint vertices sequentially. If graph is dense and all the edges have both endpoints inside the memory, all the vertices will be processed sequentially without any parallelism.

We propose a disk-based graph processing engine following asynchronous model of computation which efficiently processes billion-scale graphs on a single PC. We introduce a new processing model BiShard Parallel (BSP) based on a vertex-centric approach [2]. BSP divides the graph into several subgraphs. For every subgraph, it manages the in and out edges separately, which allows to load a subgraph with only two reads. This storage structure manages two copies of every edge (one is each direction). This setting allows every vertex to have its own copy of the edges, and ensures full parallel processing. Our contributions are two folds: 1) a new storage structure that reduces the number of I/Os significantly; and 2) a new processing model that ensures full CPU parallelism.

The rest of this paper is organized as follows. Section 2 reviews related works. Section 3 lists the core idea of the paper in details. Section 4 discusses experimental settings, results and comparison with state-of-the-art. Finally Section 5 summarizes and concludes the paper.

## 2   Related Work

GraphChi extended the work of Bender [3] and Chen et. al. [4] and proposed a mechanism that stores and processes billion-scale graphs on a single consumer PC. Their system implements a naive technique for disk-based graph processing called Parallel Sliding Window (PSW). PSW exploits the sequential I/Os and parallel computation using vertex-centric approach for processing the graph. Their results show that GraphChi out-performs all the representative disk-based distributed systems on per node basis. However, experimental results also show some bottlenecks in part of parallel graph processing and number of disk reads in every execution interval.

Recently, another disk-based graph processing framework TurboGraph [5] was proposed. TurboGraph is also designed to process very large graphs on modern consumer level machine with a flash drive. It implements a naive technique called Pin and Slide. TurboGraph fully exploits parallelism of flash disk and multi-core CPU. Their results show that their system outperforms GraphChi by an order-of-magnitude. However their solution exploits some specific properties of the flash drive which are not available on the rotational disks.

## 3   BiShard Parallel Processor

This section describes our proposed solution BiShard Parallel Processor (BPP). BPP is an asynchronous disk-based framework for processing very large graph on a single PC. We introduce a new processing model BiShard Parallel (BP). BP extends the storage structure proposed by GraphChi to reduce the non-sequential I/Os and eliminate the race conditions while accessing the edges shared by the vertices in memory. Vertices are allowed to modify their associated values and the values associated with their out-edges. BP processes the graph one chunk at a time. Processing of every interval consists of 3 steps 1) load a chunk of graph inside the memory; 2) perform computation on the chunk and modify the associated values of the vertices and out-edges; and 3) write the updated values back to the disk.

### 3.1   Loading of Subgraph

We divide the graph *G* into *P* number of intervals. Every interval *p* consists of a subset of vertices V. We associate two shards *in-shard(p)* and *out-shard(p)* with every interval *p*. in-shard(p) contains all the in-edges of the vertices in interval p, while *out-shard(p)* contains all the out-edges. In both of the shards edges are stored in order of their source vertex. Size and number of P are chosen such that any interval p can be loaded inside the memory.

In order to process the interval subgraph we need to read all its vertices and their edges from the disk. We load the in-edges of the subgraph from the *in-shard(p)* and the out-edges of the subgraph from the *out-shard(p)*. We perform only 2 non-sequential disk reads while processing an interval, irrespective of the graph size and total number of shards P.

### 3.2   Parallel vertex updates

Graph algorithms are executed on graph by defining an update function. After all the interval vertices along their in and out edges are loaded inside the memory we run the update function for all the vertices in parallel. As every vertex has its own copy of the edges, there is no race condition while accessing any of the edges. We utilize true CPU parallelism by eliminating the race conditions for accessing the edges.

### 3.3   Writing back to disk

To keep the asynchronous processing model intact we must write back the updated values to the disk as updates need to be available to any subsequent processing steps. We distribute and write the updated out-edges to all of the in-shards. Updated edges occur in sequential chunks in every shard. We keep track of the offsets in all the shards where we need to write the updates.

## 4 Experimental Evaluation

We now list our experimental settings, details of the experiments and comparison with state-of-the-art.

### 4.1 Experiments setup

Experiments were performed on a compatible PC with 3.3GHz Intel Core i5-3550 CPU, 4GB of installed main memory and a 500GB 7200 rpm disk drive. We ran Microsoft Windows 7 64-bit with default settings. We disabled the file caching to get meaningful comparisons for small and large input files.

### 4.2 Experimental results and comparisons

We implemented page rank algorithm in the same way as implemented by GraphChi. We conducted our experiments on the wiki-vote [6] data set with 7 thousand plus vertices and more than 1 hundred thousand edges. We varied the number of shards to evaluate the effect of different number of inter-interval edges and number non-sequential I/O.

Our results showed that BPP significantly outperform GraphChi with both small and large number of shards. We noticed that the performance margin is larger when number of inter-interval edges is large and performance margin gets smaller with increase in number of shards.

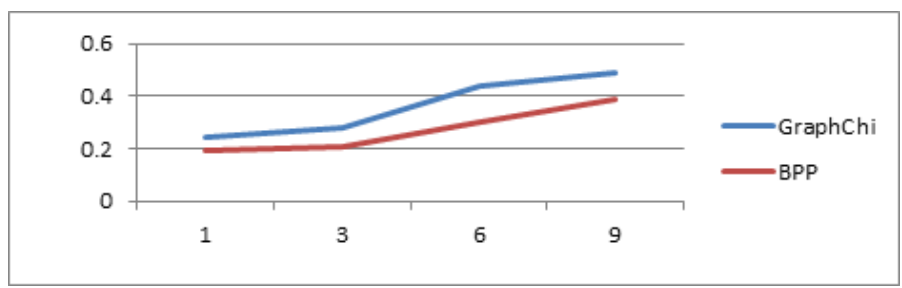

**Fig.1.** Graph visualizing the comparison of execution time of GraphChi and BPP while running Page Rank algorithm.

## 5 Conclusion

GraphChi is a single PC, disk-based, graph processing engine. It solves the problems observed in distributed processing system but suffers from serious performance issues. In this work we extend the storage structure proposed by GraphChi, and proposed a new processing method called *BiShard Parallel* (BP). We showed by theoretical analysis that our solution reduced the number of non-sequential seeks and

  

I/Os incurred in GraphChi to almost half, which makes it a better choice for processing graph on SSD. We also eliminated the race conditions between the vertices to access a common edge, which hindered GraphChi from full parallel processing of the graph vertices.

**Acknowledgements.** This research was supported by the MSIP (Ministry of Science, ICT & Future Planning), Korea, under the ITRC(Information Technology Research Center) support program (NIPA-2013- H0301-13-4006) supervised by the NIPA(National IT Industry Promotion Agency).